\documentclass[preprint,pra,a4paper,floatfix,amsmath,amssymb]{revtex4}
\usepackage{pstricks}
\usepackage{fancybox}
\usepackage{epsf}
\usepackage{times}

\newcommand{\bv}[1]{\mbox{\boldmath $#1$}}
\newcommand{\bvsub}[1]{\mbox{\scriptsize\boldmath $#1$}}
\newcommand{\intensity}[2]{$\mbox{#1}\times10^{\mbox{\scriptsize #2}}\mbox{
W/cm}^{\mbox{\scriptsize 2}}$}
   
\def\jpbo{J. Phys. B: At. Mol. Opt. Phys. }   

\def\pra{Phys. Rev. A }

\def\prl{Phys. Rev. Lett. }
\begin{document}
\title{Molecular effects in the ionization of N$_2$, O$_2$ and F$_2$ by intense laser fields}
\author{Daniel Dundas\footnote{Current Address: Max Planck Institute for Nuclear Physics,
Saupfercheckweg 1, 69117 Heidelberg, Germany} and Jan M Rost}
\affiliation{Max Planck Institute for the Physics of Complex Systems, \\
N\"{o}thnitzer Strasse 38, 01187 Dresden, Germany}
\date{\today}

\begin{abstract}
Single ionization of N$_2$, O$_2$ and F$_2$ by intense laser fields is simulated using a
time-dependent density functional approach within the exchange-only local density approximation. 
In particular, the response to laser pulses having a wavelength of $\lambda=390$nm
(corresponding to a frequency-doubled Ti:Sapphire laser) is investigated. Single ionization
yields of N$_2$ and F$_2$ are seen to agree to within a factor of 3, in keeping with experimental
results but in disagreement with predictions based on the symmetry of the highest occupied molecular orbital. The underlying mechanism for
this appears to be the importance of multi-electron molecular effects in the single ionization of 
F$_2$. In addition single ionization of O$_2$ is calculated assuming the ground state
configuration to be either a singlet or triplet state. It is discovered that the single 
ionization process is largely insensitive to the multiplicity of the initial state.
\end{abstract}
\maketitle

\begin{section}{Introduction}
In the last two decades the interaction of atoms and molecules with intense laser pulses 
has attracted widespread research. This highly non-perturbative interaction results in a number of
non-linear processes such as ionization, harmonic generation and above-threshold ionization. At the
heart of all these processes is single electron ionization which gives rise, either directly or
indirectly, to the other processes.

Compared to atoms, molecules represent a more complex class of system due to
their multi-centre
nature which introduces additional vibrational and rotational degrees of freedom. It is perhaps
surprising therefore that the ionization of molecules by intense laser pulses share many of the
characteristics with the ionization of atoms. The mechanisms of ionization can be characterized 
as either multiphoton transitions or tunnelling ionization, or some combination of both. The
process can be classified into either the tunnelling or multiphoton regime by
the Keldysh parameter, defined as $\gamma_k \equiv \sqrt{I_p/ 2U_p}$, where 
the internal binding energy is $I_p$ and the external laser-driven 
kinetic energy is $U_p$. In the long-wavelength, intense field limit $\gamma_k \ll 1$ and
tunnelling dominates and thus the excited states spectrum should play no r\^{o}le in the ionization
process. 

In a range of early experiments this was found to be the case: it was discovered that 
single ionization rates for molecules are roughly identical to those of noble gas 
atoms provided the ionization potentials were 
comparable~\cite{gibson:1991,chin:1992,walsh:1993,walsh:1994}. In one of these 
experiments~\cite{walsh:1994} the single ionization rates of O$_2$ (I$_p$ = 12.07eV) and of its 
companion noble gas atom xenon (I$_p$ = 12.13eV) were measured simultaneously by 
exposing a target  containing a mixture of xenon atoms and O$_2$ to a laser pulse 
of wavelength $\lambda =$ 10.6$\mu$m. The ionization rates were found to be similar.

However, in later experiments carried out at Ti:Sapphire laser wavelengths ($\lambda \sim$ 800nm) 
it was found that while most atom-molecules pairs obeyed this finding -- such as N$_2$ (I$_p$ =
15.58eV) and its companion atom argon (I$_p$ = 15.76eV) -- a number did not. For example,
the single ionization rate of O$_2$ was suppressed with respect to ionization 
rate of xenon by several orders of magnitude~\cite{talebpour:1996,guo:1998} while ionization of D$_2$
was also suppressed with respect to its companion noble gas atom, argon.

Considerable interest has been generated by these findings and a number of explanations have been  put forward for the origin of the suppression, most notably in O$_2$. Talebpour et 
al~\cite{talebpour:1996} suggested a
dissociative recombination process leading to a decrease in the single ionization signal. However,
later experiments~\cite{guo:1998} concluded that dissociative recombination cannot be the cause of
suppression since the suppression was observed for a range of laser polarization ellipticities. An
electronic correction to tunnelling theory was suggested by Guo~\cite{guo:2000} and reproduced the
correct suppression of O$_2$, provided that correct parameters for an effective nuclear charge and
ionization potential are chosen.

Muth-B\"{o}hm et al~\cite{muth:2000} explained the suppression of the O$_2$ single ionization 
rate using a generalization of intense-field many-body $S$-matrix theory (IMST) which included 
an interference term. They showed with their calculations that the suppression of the O$_2$ 
signal with respect to xenon, and its absence in N$_2$/Ar, is due to a symmetry induced 
dynamical effect whereby interference between ionizing wavepackets emitted from
the two distinct nuclear centres is either destructive or constructive in the low energy 
limit, depending upon the symmetry of the highest occupied molecular orbital (HOMO). 
Thus for O$_2$, in which the HOMO has an anti-bonding character (ground state configuration 
$1\sigma_g^21\sigma_u^22\sigma_g^22\sigma_u^23\sigma_g^21\pi_u^41\pi_g^2$), single
ionization should show suppression. For N$_2$, in which the HOMO has a bonding character 
(ground state configuration $1\sigma_g^21\sigma_u^22\sigma_g^22\sigma_u^21\pi_u^43\sigma_g^2$), 
single ionization should not be suppressed. In addition, according to the
symmetry argument of IMST, when suppression does occur we can write 
\begin{equation} 
\bv{k}_N\cdot\bv{R} \ll 1,
\end{equation}
where $k_N^2/2 = N\omega - U_p - I_p$ is the kinetic energy of the electron on absorbing $N$
photons, $\omega$ the laser frequency and $R$ the bond length of the molecule. Thus, if only 
the wavelength is varied, the suppression should be enhanced at shorter wavelengths. This 
explanation is consistent with the O$_2$/Xe results where suppression of O$_2$ was observed at $\lambda =$ 800nm 
but not at $\lambda =$ 10.6$\mu$m. 

Following the symmetry argument it was postulated that since F$_2$ (I$_p$ = 15.69eV) has 
an ionization potential similar to N$_2$ and argon but has valence electrons with the same symmetry
as O$_2$  (ground state configuration of F$_2$: 
$1\sigma_g^21\sigma_u^22\sigma_g^22\sigma_u^23\sigma_g^21\pi_u^41\pi_g^4$) ionization of F$_2$ 
should also be suppressed with respect to either N$_2$ or argon~\cite{muth:2000}. 
However a later experimental study~\cite{dewitt:2001} showed that the 
ionization of F$_2$ is not suppressed with respect to that of N$_2$.

In this paper we study the response in time of N$_2$, O$_2$ and F$_2$ to laser pulses
having a wavelength of 390nm and 300nm. We find  single ionization suppression in O$_2$ and its
absence in F$_2$, in accordance with the experimental results at $\lambda = 800$nm.
Within our framework of  a time-dependent density functional approach we are able to 
explain the deviations from the predictions based on the symmetry of the HOMO.
The paper is arranged as follows. In section~\ref{sec:method} the time-dependent 
density functional approach is set out and the procedure for calculating single ionization rates
described. In section~\ref{sec:results} the time-evolution will be considered. Finally, some 
conclusions will be drawn in 
section~\ref{sec:conclusions}.
\end{section}
\begin{section}{Method}
\label{sec:method}
The time-dependent density functional method provides the most detailed, practical 
and feasible ab initio approach for tackling many-body problems. Density functional 
theory (DFT), as first introduced by Hohenberg and Kohn~\cite{Hohenberg} and Kohn 
and Sham~\cite{Kohn} describes a system of interacting particles in terms of its 
density. The theory is based on the existence of an exact mapping between densities 
and external potentials and leads to the density of the interacting system being 
obtained from the density of an auxiliary system of non-interacting particles moving 
in an effective local single particle potential, i.e. the particle interactions are 
treated in an averaged-over manner. A time-dependent formalism of DFT (TDDFT) was 
provided by Runge and Gross~\cite{Runge}, who showed that the time-dependent density 
could be obtained from the response of non-interacting particles to the 
time-dependent local effective potential. In principle, many-body effects are 
included exactly through an exchange-correlation functional; in practice, the 
form of this functional is unknown and at best it can only be approximated.  

Such an approach has been widely used in treating the interaction of molecules with 
intense, short-duration laser pulses. For instance, in the approach of Chu and 
co-workers~\cite{chu1,chu2} a pseudo-spectral mesh technique is used in the solution of the 
Kohn-Sham equations. A non-adiabatic quantum molecular dynamics (NA-QMD) method, in which the 
electron dynamics are treated quantum mechanically using TDDFT, has been developed by
Schmidt and co-workers~\cite{Uhlmann}. In this work the electron orbitals are expanded in
terms of Gaussian basis functions. More recently, Castro et al~\cite{castro:2004} developed
grid-based NA-QMD technique to study harmonic generation in H$_2$. However, these calculations 
were limited to a 1D treatment of the electron dynamics. Another grid-based DFT method has
been developed by Otobe et al~\cite{otobe:2004}. While this method has calculated tunnel
ionization rates for N$_2$, O$_2$ and F$_2$ two approximations have been made. Firstly, a
pseudo-potential description of the electron-nuclear potential was used meaning that only  the
valence electrons in the molecules were treated. Secondly, only static field ionization rates were
calculated.

The current approach utilizes a NA-QMD approach in which the electron dynamics are described by a
hybrid finite difference-Lagrange mesh technique~\cite{dundas:2004}. For the present calculations, however, a
fixed nuclei description has been employed. Details of the approach are now given. 

\begin{subsection}{Time-dependent Kohn-Sham equations}
In the TDDFT approach, the total $N_e$-electron Kohn-Sham wavefunction is written as a single 
determinant of one-particle Kohn-Sham orbitals. Denoting the spin state of each orbital by 
the label $\sigma = \uparrow, \downarrow$, we can write the electron density as  
\begin{equation}
n(\bv{r}, t) = \sum_{\sigma = \uparrow,\downarrow} n_\sigma(\bv{r}, t) =
\sum_{\sigma = \uparrow,\downarrow} 
\sum_{i=1}^{N\sigma} \left|\psi_{i\sigma}(\bv{r}, t)\right|^2,
\end{equation}
where $N_\sigma$ is the number of electrons in spin state $\sigma$ and $\psi_{i\sigma}(\bv{r}, t)$
are the Kohn-Sham orbitals.
The orbitals are obtained through the solution of the time-dependent Kohn-Sham equations
\begin{equation}
\label{eq:kohn-sham-equations}
i\frac{\partial}{\partial t} \psi_{i\sigma}(\bv{r}, t) = 
\left[-\frac{1}{2}\nabla^2 + 
V_{\mbox{\scriptsize eff}}(\bv{r}, t)\right]\psi_{i\sigma}(\bv{r}, t),
\end{equation}
where  
\begin{equation}
V_{\mbox{\scriptsize eff}}(\bv{r}, t) = 
V_{\mbox{\scriptsize ext}}(\bv{r}, \bv{R}, t) + V_{H}(\bv{r}, t) 
+ V_{xc\sigma}(\bv{r}, t), 
\end{equation}
is the time-dependent effective potential which is given in terms of the external
potential 
\begin{equation}
V_{\mbox{\scriptsize ext}}(\bv{r}_i, \bv{R}, t) = 
V_{\mbox{\scriptsize ions}}(\bv{r}_i, \bv{R}, t) + 
U_{\mbox{\scriptsize elec}}(\bv{r}_i, t),
\label{eq:pot_ext}
\end{equation}
where $U_{\mbox{\scriptsize elec}}(\bv{r}_i, t)$ denotes the interaction between
electron $i$ and the applied laser field and where
\begin{equation}
\label{eq:ion_pot}
V_{\mbox{\scriptsize ions}}(\bv{r}_i, \bv{R}, t) = \sum_{I=1}^{N_n} V_{\mbox{\scriptsize
ion}}(\bv{r}_i, \bv{R}_I, t) =
-\sum_{I=1}^{N_n}\frac{Z_I}{\left|\bv{R}_I - \bv{r}_i\right|},
\end{equation}
denotes the Coulomb interaction between electron $i$ and all ions. The time-dependent 
effective potential also depends on the Hartree potential 
\begin{equation}
V_{H}(\bv{r}, t) = \int d\bv{r}^\prime \frac{n(\bv{r}^\prime, t)}{\left|\bv{r} -
\bv{r}^\prime\right|},
\label{eq:hartree}
\end{equation}
and the exchange-correlation potential 
\begin{equation}
V_{xc\sigma}(\bv{r}, t) = \left.\frac{\delta E_{xc}\left[ n_\uparrow,
n_\downarrow\right]}{\delta n_\sigma}\right|_{n_\sigma =
n_\sigma(\bvsub{r}, t)},
\end{equation}
where $E_{xc}\left[ n_\uparrow, n_\downarrow\right]$ is the exchange-correlation
action. 
\end{subsection}
\begin{subsection}{Treatment of the exchange-correlation potential}
All many-body effects are included within the
exchange-correlation potential, which in practice must be approximated. While many
sophisticated approximations to this potential have been 
developed~\cite{functionals}, the simplest is the adiabatic local density 
approximation in the exchange-only limit (xLDA). In this case the exchange 
energy functional is given by 
\begin{equation}
	E_x\left[ n_\uparrow, n_\downarrow\right] = 
	-\frac{3}{2} \left(\frac{3}{4\pi}\right)^{1/3}
	          \sum_{\sigma=\uparrow,\downarrow}\int d\bv{r}\, 
		  n_\sigma^{4/3}(\bv{r}, t),
\end{equation}
from which the exchange-correlation potential
\begin{equation}
	V_{xc\sigma}(\bv{r}, t) = - \left(\frac{6}{\pi}\right)^{1/3}
	          n_\sigma^{1/3}(\bv{r}, t),
\end{equation}
can be obtained. This approximate functional is  easy to implement
and is one of the most widely used exchange-correlation functionals. However, it
does suffer from a number of drawbacks, most notably it contains self-interaction
errors. This self-interaction means that the asymptotic form of the potential is 
exponential instead of Coulombic. Therefore, the electronic properties and 
response of the system can differ markedly from those of the actual system (as we
shall see, for example in section~\ref{sec:ionization_potential}).
\end{subsection}
\begin{subsection}{Numerical details}

Precise numerical details of how the code is implemented are given in~\cite{dundas:2004}.
Briefly, the numerical implementation of TDDFT uses a cylindrical grid treatment of the 
electronic Kohn-Sham orbitals. As in~\cite{Dundas2} a finite difference treatment of 
the $z$-coordinate and a Lagrange mesh treatment of the $\rho$-coordinate based upon 
Laguerre polynomials is used. For the case of diatomic molecules and considering a linearly polarized
laser pulse with the laser polarization direction parallel to the molecular axis the azimuthal 
angle $\phi$ can be treated analytically. The time-dependent Kohn-Sham equations of TDDFT are 
discretized in space using these grid techniques and the resulting computer code 
parallelized to run on massively-parallel processors. Several parameters in the code affect the
accuracy of the method. These are the number of points in the finite difference grid ($N_z$), the
finite difference grid spacing ($\Delta z$), the number of Lagrange-Laguerre mesh points ($N_\rho$), 
the scaling parameter of the Lagrange-Laguerre mesh ($h_\rho$), the order of the time propagator 
($N_t$) and the time spacing ($\Delta t$)~\cite{dundas:2004}. In all the calculations presented 
here, converged results were obtained using the following parameters (atomic units are used
throughout):  $N_z = 2291$, $\Delta z = 0.05$, $N_\rho = 43$, $h_\rho = 0.28838771$, 
$N_t = 18$ and $\Delta t = 0.02$. The code was parallelized and the calculations were 
carried out using 79 processors.
\end{subsection}
\begin{subsection}{Determination of Ionization}

Within TDDFT all observables are functionals of the electronic density. However, as in the case of the
exchange correlation functional, the exact forms of these functionals may not be known and must therefore be
approximated. The functional describing ionization falls into this category. To date, most calculations of
ionization within TDDFT have been obtained using geometric properties of the time-dependent Kohn-Sham
orbitals~\cite{ullrich:2000}. Briefly, an analysing box is introduced as a way to approximately separate
the bound- and continuum-state parts of the wavefunction. We define the box such that all relevant bound
states of the wavefunction are contained inside this box while the continuum contributions are found
outside the box. In that case the number of bound electrons is given by
\begin{equation}
N_{\mbox{\scriptsize bound}}(t) = \int_{\mbox{\scriptsize inside box}} d\bv{r}\,\, n(\bv{r}, t),
\end{equation}
while the number of continuum electrons is given by
\begin{equation}
N_{\mbox{\scriptsize esc}}(t) = 
\int_{\mbox{\scriptsize outside box}} 
d\bv{r}\,\, n(\bv{r}, t).
\end{equation}
We then define the number of bound and continuum electrons for each Kohn-Sham orbital as (dropping spin
subscripts)
\begin{equation}
N_j(t) = \int_{\mbox{\scriptsize inside box}} d\bv{r}\,\, \left|\psi_j(\bv{r}, t)\right|^2,
\end{equation}
for bound electrons and
\begin{equation}
\overline{N}_j(t) = \int_{\mbox{\scriptsize outside box}} d\bv{r}\,\, \left|\psi_j(\bv{r}, t)\right|^2,
\end{equation}
for continuum electrons. Thus we obtain approximate ion probabilities, $P^k(t)$, for charge state $k$. In
particular we find
\begin{equation}
P^0(t) = N_1(t) \cdots N_{N_e}(t),
\end{equation}
and
\begin{equation}
P^1(t) = \sum_{n=1}^{N_e} N_1(t) \cdots \overline{N}_n(t) \cdots N_{N_e}(t).
\end{equation}

\end{subsection}

\end{section}
\begin{section}{Results}
\label{sec:results}
We now study the response of N$_2$, O$_2$ and F$_2$ to intense laser pulses. Firstly, we
compute the ionization potentials of the various molecules. Then we compare single
ionization of N$_2$ and F$_2$ showing that the F$_2^+$ yield shows no suppression with
respect to the N$_2^+$ yield. Thirdly, we compare the yields of O$_2^+$ using two initial
configurations of the molecule, namely a singlet and a triplet state. This shows that the
suppression of the O$_2^+$ signal is not due to the fact that the O$_2$ ground state is a triplet
configuration. In order to gain a better understanding of the electronic dynamics we analyse
the time-evolution of the Kohn-Sham orbitals, showing that in the case of F$_2$ the response
of the core electrons show a significant response to the pulse.

\begin{subsection}{Ionization Potentials}

\label{sec:ionization_potential}
The starting point for any simulation of the response of diatomic molecules to
intense laser pulses is an appropriate description of the field-free structure.
In our calculations we assume that the molecules are initially in their ground
states. Starting from the equilibrium ion separation the ground state electronic 
density is calculated self-consistently from the time-dependent Kohn-Sham 
equations using an iterative Lanzcos method~\cite{smyth:1998,Dundas2}. Of particular importance
in the present calculations are the ionization potentials of the different 
molecules. These are obtained by first calculating the ground state energy of
the particular molecule and then that of the molecule with one
electron removed. The results are presented in table~\ref{tab:ionpot} for N$_2$, 
O$_2$ and F$_2$. We see that the ionization potentials of N$_2$ and O$_2$ compare well to 
the experimental values~\cite{hertz} while the ionization potential of F$_2$ shows a 10\% 
difference with the experimental value. The difference can be explained in terms of the 
choice of exchange-correlation potential, as discussed earlier. Obviously in
comparing single ionization in N$_2$ and F$_2$ the errors 
in their ionization potentials will have an important impact upon their 
subsequent response. It has been pointed out to us~\cite{becker:2004} that the
10\% decrease in the ionization potential of F$_2$ could increase the single ionization yield by two
orders of magnitude. However, such an estimation, obtained using IMST, only takes into account the 
response of the HOMO to the field. As our results will show,
it appears that molecular effects due to the other orbitals are extremely important in this system
and  such a large increase in ionization yield due to a lowering of
the ionization potential is unlikely. This is supported by the static
field tunnel ionization calculations of 
Otobe et al~\cite{otobe:2004} who employed a DFT approach using the self-interaction free KLI approximation. 
In these calculation the ionization potential of F$_2$ was much closer to the experimental value. However, 
no suppression of the F$_2^+$ signal was evident.

\end{subsection}
\begin{subsection}{Single ionization yields of N$_2$ and F$_2$}

Table~\ref{tab:n2f2_ion1} presents a comparison of the
single ionization yields for N$_2^+$ and F$_2^+$ after the interaction of the neutral molecules with
a 24 cycle laser pulse (pulse length, $\tau =$ 31.2fs) 
having a wavelength of $\lambda =$ 390nm over the intensity range $I=$
\intensity{1}{14} to $I=$ \intensity{8}{14}. As discussed in the introduction, if the symmetry
induced dynamical effect is responsible for ionization suppression, equation~(1) predicts
that the yield of F$_2^+$ ions should be suppressed to a greater extent to the yield of 
N$_2^+$ ions at the wavelength used in these calculations. While differences do exist between the
N$_2^+$ and F$_2^+$ ion yields, we see that the differences are no more than a factor of three. 
We conclude that {\em single ionization of F$_2$ is not suppressed with respect to the single
ionization of N$_2$}. 

\end{subsection}
\begin{subsection}{Single ionization yields from $^1$O$_2$ and $^3$O$_2$}

One explanation postulated for the nonatomiclike single ionization in 
O$_2$ compared to single ionization of N$_2$ is the fact that the ground state
of O$_2$ is a triplet state with a half-filled open-shell structure whereas the 
ground state of N$_2$ is a singlet state with a closed-shell structure~\cite{guo:1998}.
M\"{u}th-Bohm et al~\cite{muth:2000} argued that such a difference in the spin states is unlikely to
influence the ionization signal since the spin-degrees of freedom are not effectively coupled to a
dipole field. We are unaware of any calculation to date in which single ionization of O$_2$ is studied
as a function of the multiplicity of the ground state.

Table~\ref{tab:sing_trip} compares the single ionization yields of $^1$O$_2$ and 
$^3$O$_2$ at the end of a 24 cycle laser pulse ($\tau =$ 31.2fs) having a wavelength 
of $\lambda =$ 390nm over the intensity range $I=$ \intensity{1}{14} to $I=$ \intensity{8}{14}.  
We see that no significant difference of the yields due to the two configurations is evident.
Hence, we conclude that {\em the suppression of O$_2^+$ with respect to that of Xe$^+$ is not 
due to the ground state of O$_2$ being a triplet state}.

\end{subsection}
\begin{subsection}{Orbital response of N$_2$, O$_2$ and F$_2$}

In order to gain a better understanding of the ionization dynamics we have investigated the time 
evolution of the Kohn-Sham orbitals. It must be stressed that these orbitals do not have
any physical significance with the molecular orbitals of the actual system. However,
studying the evolution of the Kohn-Sham orbitals allows us to obtain information about
the orbital symmetries. In figure~\ref{fig:n2_intensity} we present results for the time-evolution
of the Kohn-Sham orbitals of N$_2$ during its interaction with a 24 cycle laser pulse of
wavelength $\lambda =$ 390nm. Two laser intensities are presented: $I=$ \intensity{1}{14} and 
$I=$ \intensity{6}{14}. It can clearly be seen that the orbital having the same symmetry ($3\sigma_g$) as 
the valence orbital of N$_2$ shows the dominant response to the field. The same behaviour is
observed for all other laser intensities considered in table~\ref{tab:n2f2_ion1}. Thus it is
apparent that single ionization of N$_2$ occurs predominantly by ionization of the valence
electron. Since the valence orbital of N$_2$ has a bonding character then,
based upon the symmetry arguments of IMST, we would expect no ionization suppression to occur -- 
as is the case.

In figure~\ref{fig:o2_intensity} we present results for the time-evolution of the Kohn-Sham 
orbitals of O$_2$ during its interaction with a 24 cycle laser pulse of
wavelength $\lambda =$ 390nm. Again, two laser intensities are presented: $I=$ \intensity{1}{14} and 
$I=$ \intensity{6}{14}. As in the case of N$_2$ we see that the orbital having the same 
symmetry ($1\pi_g$) as the valence orbital of O$_2$ shows the dominant response to the field
and thus single ionization of O$_2$ occurs predominantly by ionization of the valence
electron. Since the valence orbital of O$_2$ has an anti-bonding character then,
based upon the symmetry argument of IMST, we would expect ionization suppression to 
occur -- again, in accordance with the numerical results and experiment.

In figure~\ref{fig:f2_intensity} we present results for the time-evolution of the Kohn-Sham 
orbitals of F$_2$ during its interaction with a 24 cycle laser pulse of
wavelength $\lambda =$ 390nm. Four laser intensities are presented: $I=$ \intensity{1}{14}, 
$I=$ \intensity{2}{14}, $I=$ \intensity{4}{14} and $I=$ \intensity{6}{14}. 
It is apparent that the orbital having the same symmetry ($1\pi_g$) as 
the valence orbital of F$_2$ does not in this case show the dominant response 
to the field. Instead we see that for all of the intensities, apart from $I=$ \intensity{1}{14},
the orbital having the symmetry $3\sigma_g$ shows the dominant response. Since this orbital
has a bonding character then, based upon the predictions of IMST, ionization suppression will 
not occur. We conclude that single ionization of F$_2$ does not occur predominantly by 
ionization of the valence electron. Thus, molecular structure effects for this molecule 
are of crucial importance.

In understanding why the orbital having the same symmetry ($1\pi_g$) as the valence orbital 
of F$_2$ does not respond predominantly to the field we consider the response of F$_2$ at 
the laser intensity of $I =$ \intensity{1}{14} where the $3\sigma_g$ orbital did not 
show the dominant response to the field. In this case the $1\pi_g$ anti-bonding orbital and the 
$1\pi_u$ bonding orbital showed the dominant response. Indeed we see that over all laser 
intensities these two orbitals response in a similar fashion to the field. It would 
therefore appear that an interference is occurring between these two orbitals. 

To gain further insight we have repeated   the TDDFT calculations 
at the laser intensity of $I =$ \intensity{2}{14}, whereby only a subset of the 
orbitals were allowed to respond to the field, all other orbitals being frozen. 
The results are presented in figure~\ref{fig:f2_orbits}. We see from 
figure~\ref{fig:f2_orbits}(c) that when only the 
$1\pi_u$ and $1\pi_g$ orbitals respond to the field, ionization is suppressed. In 
figures~\ref{fig:f2_orbits}(a) and~\ref{fig:f2_orbits}(b) we see that the ionization 
of either the $1\pi_u$ or $1\pi_g$ is greater than
that of the $3\sigma_g$ orbital. In figure~\ref{fig:f2_orbits}(d), when the 
$3\sigma_g$, $1\pi_u$ and $1\pi_g$ orbitals respond, ionization of the $3\sigma_g$ 
is dominant. This behavior illustrates a rather complicated correlated response of the orbitals to the laser field.
Moreover,  from figure~\ref{fig:f2_intensity} we can see that a resonance effect is coming into play. This
is evidenced at the laser intensities of $I =$ \intensity{4}{14} and $I =$ \intensity{6}{14} where
the rate of depletion of the $1\pi_u$ and $1\pi_g$ orbitals is not exponential but instead
shows evidence of two distinct rates being present, namely an intermediate resonance state is initially
populated and then undergoes ionization itself. Figure~\ref{fig:f2_intensity_300nm} presents 
results for the time-evolution of the Kohn-Sham orbitals of F$_2$ during its interaction with a 
24 cycle laser pulse of wavelength $\lambda =$ 300nm at laser intensities $I=$ \intensity{4}{14} 
and $I=$ \intensity{6}{14}. In this case the orbital populations are depleted exponentially, thus confirming
the presence of the resonance at $\lambda =$ 390nm.

Hence, we confirm   the connection of the suppression of ionization with
destructive interference of outgoing electron waves from the ionized electron orbital as put forward first by
Muth-B\"ohm et al.\ \cite{muth:2000}. However, the prediction of ionization suppression justified within the IMST approach
through the symmetry of the HOMO is not reliable, since it turns out that, e.g., in the case of F$_2$ the electronic
response to the laser pulse is rather complicated and does not lead to dominant depletion of the HOMO. Therefore,
the symmetry of the HOMO is not sufficient to predict ionization suppression. However, at 
least for F$_2$, the symmetry of the dominantly ionized orbital is consistent with the 
non-suppression of ionization.
\end{subsection}
\end{section}

\begin{section}{Summary}
\label{sec:conclusions}
Single ionization of N$_2$, O$_2$ and F$_2$ by intense laser fields having a wavelength of 
$\lambda=390$nm has been simulated using a time-dependent density functional approach 
within the exchange-only local density approximation. The results obtained for the single
ionization of N$_2$ and O$_2$ are in agreement with IMST results~\cite{muth:2000}, if
the HOMO is dominantly ionized. This is the case for N$_2$ and O$_2$, where for the latter
molecule destructive interference
leads to a suppression of ionization. In addition it is found that the suppressed ionization of
O$_2$ with respect to its companion noble gas atom (xenon) does not arise from the multiplicity of
the ground state of O$_2$. 

For single ionization of F$_2$, it is found that the ionization signal
is not suppressed with respect to the N$_2^+$ yield, in keeping with other static-field TDDFT results~\cite{otobe:2004}.
Multielectron molecular correlation leads to the dominant ionization of an orbital with symmetry different from the one of the HOMO.
This is the reason why the prediction based on the symmetry of the HOMO fails \cite{muth:2000}.

The present calculations are carried out within xLDA.  Implementations of TDDFT which go beyond the xLDA approximation to
a self-interaction free exchange-correlation potential  improve the description of the
ionization potentials and will be the focus of future work. In
addition only parallel transition have been considered in this paper which is a reasonable approximation for strong field ionization of diatomics. Future work will focus on generalizing 
the calculations of single ionziation rates to arbitrary orientations between the molecular axis
and the laser polarization direction.

\end{section}
\begin{acknowledgments}
The authors would like to acknowledge useful discussions with Andreas Becker.
The work reported in this paper is supported by the UK Engineering and Physical Sciences 
Research Council by provision computer resources at Computer Services for Academic 
Research, University of Manchester and HPC(X), Daresbury Laboratory. 
\end{acknowledgments}

\newpage
\begin{table}[p]
\begin{center}
\begin{tabular}{@{}ccccc}
\hline
 &\multicolumn{3}{c}{Ionization Potential (eV)} \\   
\hline
  & N$_2$ & O$_2$ & F$_2$ \\
\hline    
 Present    &  15.91 &  11.45 &  14.14 \\
 Exact      &  15.58 &  12.07 &  15.69\\
 \hline 
\end{tabular}
\end{center}
\caption{Ionization potentials of N$_2$, O$_2$ and F$_2$ calculated using a TDDFT
approach within the exchange-only local density approximation compared with the 
experimental values~\cite{hertz}. While the ionization potentials of N$_2$
and O$_2$ agree to within 5\%, those for F$_2$ differ by 10\%.}
\label{tab:ionpot}
\end{table}
\newpage
\begin{table}[p]
\begin{center}
\begin{tabular}{@{}ccccc}
\hline
Laser Intensity & \mbox{\hspace*{0.2cm}} & \multicolumn{3}{c}{Ion Yields} \\
(\intensity{}{14}) & \mbox{} & N$_2^+$ & \mbox{\hspace*{0.2cm}} & F$_2^+$ \\
\hline    
 1.0 & \mbox{} & 0.2479176 & \mbox{} & 0.1051454 \\
 2.0 & \mbox{} & 0.3804471 & \mbox{} & 0.2523075 \\
 4.0 & \mbox{} & 0.4025715 & \mbox{} & 0.3742050 \\
 6.0 & \mbox{} & 0.3501940 & \mbox{} & 0.3278126 \\
 8.0 & \mbox{} & 0.2919808 & \mbox{} & 0.3375370 \\
 \hline 
\end{tabular}
\end{center}
\caption{Ion yields of N$_2^+$ and F$_2^+$ after interaction of the neutral molecules with 
a 24 cycle laser pulse having a wavelength of $\lambda =$ 390nm for a range of laser 
intensities. The ion yields differ by factor of less than 3. Hence, the F$_2^+$ yield is not suppressed with
respect to the N$_2^+$ yield.}
\label{tab:n2f2_ion1}
\end{table}
\begin{table}[p]
\begin{center}
\begin{tabular}{@{}ccccc}
\hline
Laser Intensity & \mbox{\hspace*{0.2cm}} & \multicolumn{3}{c}{Ion Yield} \\
(\intensity{}{14}) & \mbox{} & $^1$O$_2 \rightarrow$ O$_2^+$ & \mbox{\hspace*{0.2cm}} & $^3$O$_2
\rightarrow$ O$_2^+$ \\
\hline    
 1.0 & \mbox{} & 0.0009385 & \mbox{} & 0.0006335 \\
 2.0 & \mbox{} & 0.0022583 & \mbox{} & 0.0017794 \\
 4.0 & \mbox{} & 0.0038756 & \mbox{} & 0.0032043 \\
 6.0 & \mbox{} & 0.0041731 & \mbox{} & 0.0033902 \\
 8.0 & \mbox{} & 0.0036198 & \mbox{} & 0.0027161 \\
 \hline 
\end{tabular}
\end{center}
\caption{Ion yields of O$_2^+$ (assuming either an initial singlet or triplet configuration of
the neutral O$_2$ molecule) after interaction with 
a 24 cycle laser pulse having a wavelength of $\lambda =$ 390nm for a range of laser 
intensities. The ion yields roughly agree, indicating that the multiplicity of the ground state of 
O$_2$ do not play a r\^{o}le in the suppression of the O$_2^+$ yield.}
\label{tab:sing_trip}
\end{table}
\begin{figure*}[p]
\epsfxsize=18cm\epsfclipon\epsffile{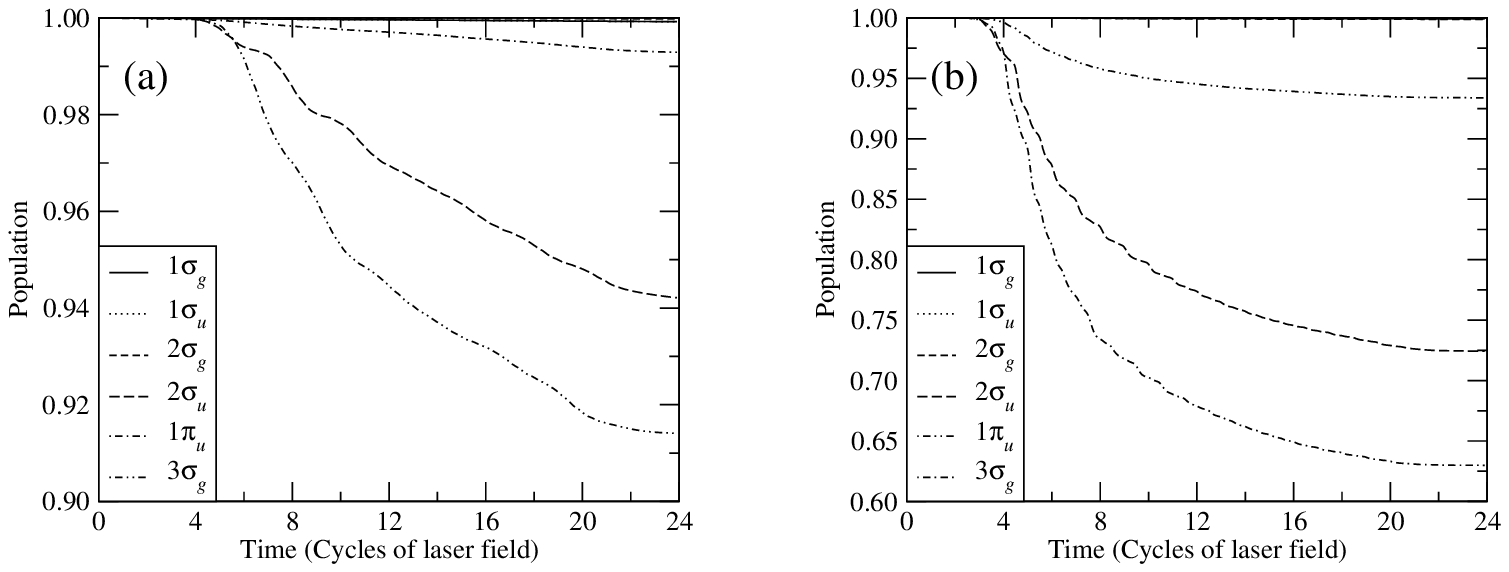}

\caption{Kohn-Sham orbital populations for N$_2$ during interaction with a 24 cycle laser pulse
having a wavelength of $\lambda =$ 390nm and laser intensity (a) $I =$ \intensity{1}{14} and 
(b) $I =$ \intensity{6}{14}. For all laser intensities
it is seen that the Kohn-Sham orbital having the same symmetry ($3\sigma_g$) as the valence orbital of N$_2$
shows the predominant response to the field.}
\label{fig:n2_intensity}
\end{figure*}
\begin{figure*}[p]
\epsfxsize=18cm\epsfclipon\epsffile{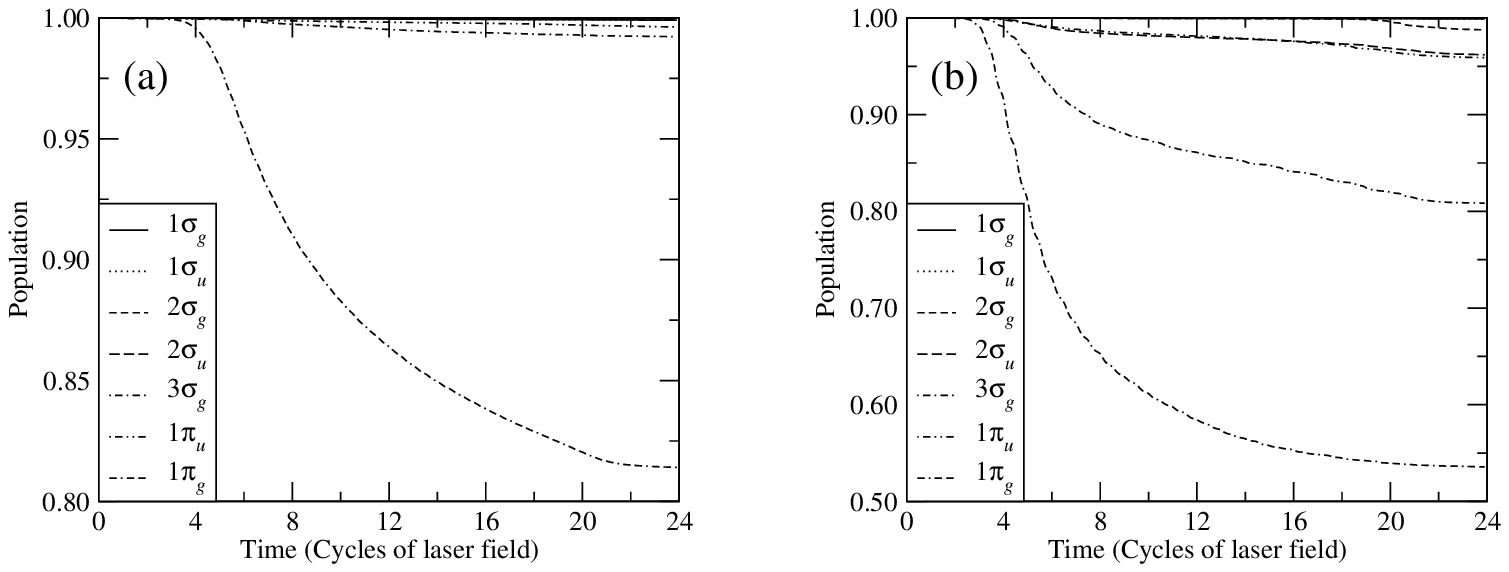}

\caption{Kohn-Sham orbital populations for O$_2$ during interaction with a 24 cycle laser pulse
having a wavelength of $\lambda =$ 390nm and laser intensity (a) $I =$ \intensity{1}{14} and 
(b) $I =$ \intensity{6}{14}. For all laser intensities
it is seen that the Kohn-Sham orbital having the same symmetry ($1\pi_g$) as the valence orbital of O$_2$
shows the predominant response to the field.}
\label{fig:o2_intensity}
\end{figure*}
\newpage
\begin{figure*}[p]
\epsfxsize=18cm\epsfclipon\epsffile{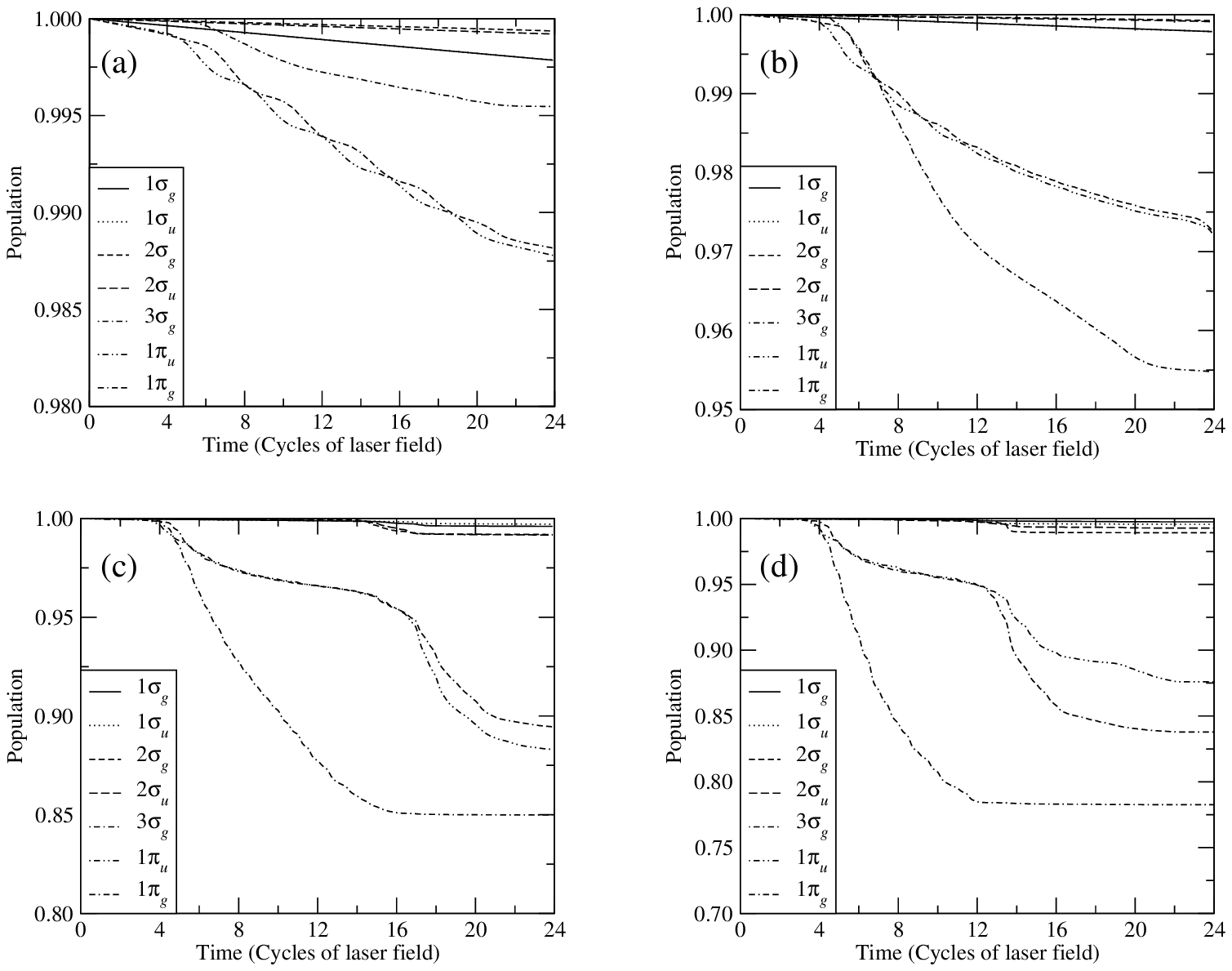}

\caption{Kohn-Sham orbital populations for F$_2$ during interaction with a 24 cycle laser pulse
having a wavelength of $\lambda =$ 390nm and laser intensity (a) $I =$ \intensity{1}{14}, 
(b) $I =$ \intensity{2}{14}, (c) $I =$ \intensity{4}{14} and (d) $I =$ \intensity{6}{14}. For all laser intensities,
except $I =$ \intensity{1}{14}, it is seen that the $3\sigma_g$ orbital predominantly responds to the
field. At the lowest intensity the $1\pi_g$ anti-bonding orbital (having the same symmetry as the
valence orbital of F$_2$) is equally dominant with the  $1\pi_u$ bonding orbital. These two orbitals
respond almost identically at all laser intensities.}
\label{fig:f2_intensity}
\end{figure*}
\begin{figure*}[p]
\epsfxsize=18cm\epsfclipon\epsffile{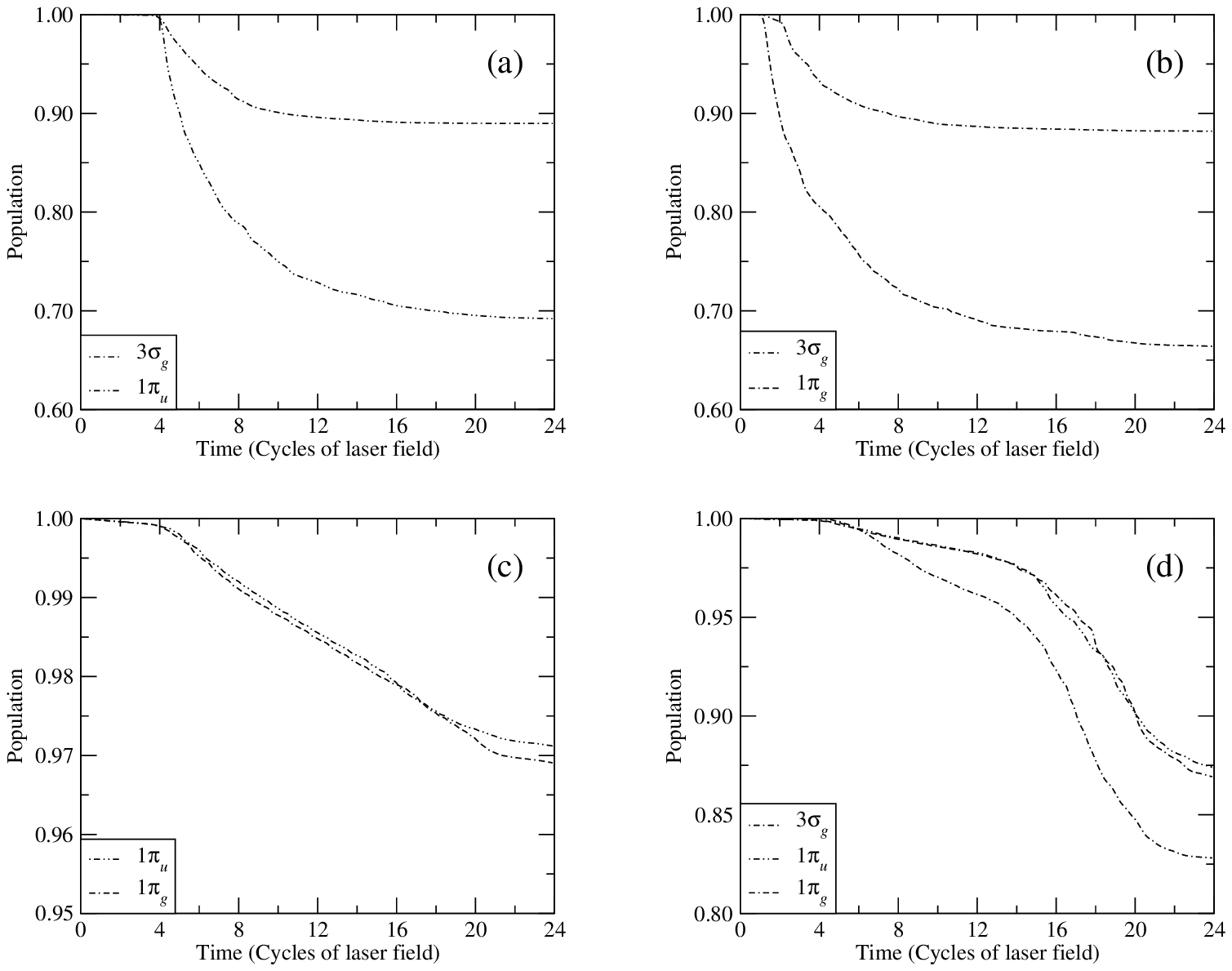}

\caption{Kohn-Sham orbital populations for F$_2$ during interaction with a 24 cycle laser pulse
having a wavelength of $\lambda =$ 390nm and an intensity of $I =$ \intensity{2}{14}. 
In these calculations only the orbitals shown in each figure respond to the field. All
other orbitals are frozen. It can clearly be seen in (c) that when only the 
$1\pi_u$ and $1\pi_g$ orbitals respond to the field, ionization is suppressed. In (a)
and (b) we see that the ionization of either the $1\pi_u$ or $1\pi_g$ is greater than
that of the $3\sigma_g$ orbital. When the $3\sigma_g$, $1\pi_u$ and $1\pi_g$ orbitals
respond, however, ionization of the $3\sigma_g$ orbital is dominant.}
\label{fig:f2_orbits}
\end{figure*}
\begin{figure*}[p]
\epsfxsize=18cm\epsfclipon\epsffile{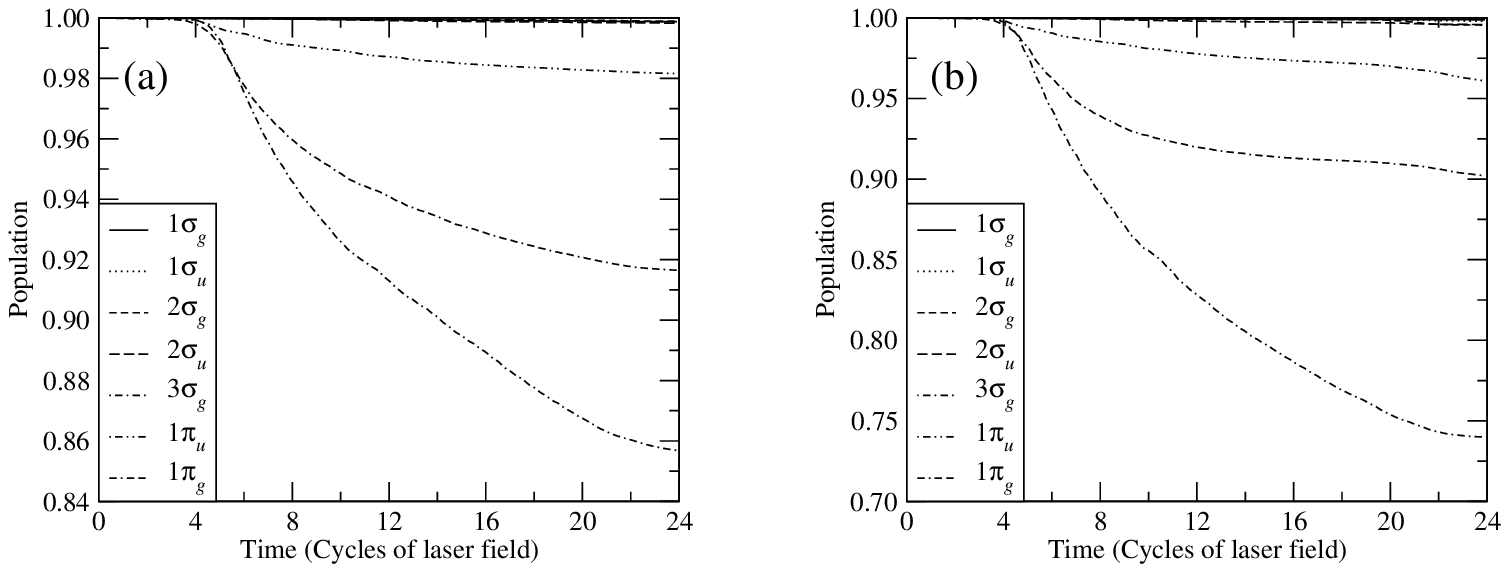}

\caption{Kohn-Sham orbital populations for F$_2$ during interaction with a 24 cycle laser pulse
having a wavelength of $\lambda =$ 300nm and intensity (a) $I =$ \intensity{4}{14} and 
(b) $I =$ \intensity{6}{14}. At this laser wavelength we see
that the the orbital populations fall off exponentially, unlike the response at $\lambda =$ 390nm observed 
in figure~\protect\ref{fig:f2_intensity}. We conclude that the response at $\lambda =$ 390nm is due to the 
population of an intermediate resonance state.}
\label{fig:f2_intensity_300nm}
\end{figure*}
\end{document}